\documentclass
[twocolumn,showpacs,preprintnumbers,amsmath,amssymb]
{revtex4}
\usepackage{graphicx}
\usepackage{dcolumn}
\usepackage{bm}
\usepackage{here}
\usepackage{color}
\usepackage{braket}
\usepackage{amsmath}  
\usepackage{amssymb}  
\usepackage{fancybox}

\begin{document}

\newcommand{\vc}[1]{\mbox{\boldmath $#1$}}
\newcommand{\fracd}[2]{\frac{\displaystyle #1}{\displaystyle #2}}
\newcommand{\red}[1]{\textcolor{red}{#1}}
\newcommand{\blue}[1]{\textcolor{blue}{#1}}
\newcommand{\green}[1]{\textcolor{green}{#1}}



\def\ni{\noindent}
\def\nn{\nonumber}
\def\bH{\begin{Huge}}
\def\eH{\end{Huge}}
\def\bL{\begin{Large}}
\def\eL{\end{Large}}
\def\bl{\begin{large}}
\def\el{\end{large}}
\def\beq{\begin{eqnarray}}
\def\eeq{\end{eqnarray}}
\def\beqnn{\begin{eqnarray*}}
\def\eeqnn{\end{eqnarray*}}

\def\bit{\begin{itemize}}
\def\eit{\end{itemize}}
\def\bsc{\begin{screen}}
\def\esc{\end{screen}}

\def\eps{\epsilon}
\def\th{\theta}
\def\del{\delta}
\def\omg{\omega}

\def\e{{\rm e}}
\def\exp{{\rm exp}}
\def\arg{{\rm arg}}
\def\Im{{\rm Im}}
\def\Re{{\rm Re}}

\def\sup{\supset}
\def\sub{\subset}
\def\a{\cap}
\def\u{\cup}
\def\bks{\backslash}

\def\ovl{\overline}
\def\unl{\underline}

\def\rar{\rightarrow}
\def\Rar{\Rightarrow}
\def\lar{\leftarrow}
\def\Lar{\Leftarrow}
\def\bar{\leftrightarrow}
\def\Bar{\Leftrightarrow}

\def\pr{\partial}

\def\>{\rangle} 
\def\<{\langle} 
\def\RR {\rangle\!\rangle} 
\def\LL {\langle\!\langle} 
\def\const{{\rm const.}}

\def\e{{\rm e}}

\def\Bstar{\bL $\star$ \eL}

\def\etath{\eta_{th}}
\def\irrev{{\mathcal R}}
\def\e{{\rm e}}
\def\noise{n}
\def\hatp{\hat{p}}
\def\hatq{\hat{q}}
\def\hatU{\hat{U}}

\def\hatA{\hat{A}}
\def\hatB{\hat{B}}
\def\hatC{\hat{C}}
\def\hatJ{\hat{J}}
\def\hatI{\hat{I}}
\def\hatP{\hat{P}}
\def\hatQ{\hat{Q}}
\def\hatU{\hat{U}}
\def\hatW{\hat{W}}
\def\hatX{\hat{X}}
\def\hatY{\hat{Y}}
\def\hatV{\hat{V}}
\def\hatt{\hat{t}}
\def\hatw{\hat{w}}

\def\hatp{\hat{p}}
\def\hatq{\hat{q}}
\def\hatU{\hat{U}}
\def\hatn{\hat{n}}

\def\hatphi{\hat{\phi}}
\def\hattheta{\hat{\theta}}

\def\iset{\mathcal{I}}
\def\fset{\mathcal{F}}
\def\pr{\partial}
\def\traj{\ell}
\def\eps{\epsilon}
\def\U{\hat{U}}

\def\U{U_{\rm cls}}
\def\P{P_{{\rm cls},\eta}}
\def\traj{\ell}
\def\cc{\cdot}

\def\DZ{D^{(0)}}
\def\Dcls{D_{\rm cls}}

\newcommand{\relmiddle}[1]{\mathrel{}\middle#1\mathrel{}}

\title{Presence and Abesnce of Delocalization-localization Transition in Coherently Perturbed
Disordered Lattices}
\author{Hiroaki S. Yamada}
\affiliation{Yamada Physics Research Laboratory,
Aoyama 5-7-14-205, Niigata 950-2002, Japan}
\author{Kensuke S. Ikeda}
\affiliation{College of Science and Engineering, Ritsumeikan University, 
Noji-higashi 1-1-1, Kusatsu 525-8577, Japan}

\date{\today}
\begin{abstract}
A new type of delocalization induced by coherent harmonic perturbations in 
one-dimensional Anderson-localized disordered systems is investigated. With only a 
few $M$ frequencies a normal diffusion is realized, but the transition to localized state 
always occurs as the perturbation strength is weakened below a critical value. 
The nature of the transition qualitatively follows the Anderson transition (AT) 
if the number of degrees of freedom $M+1$ is regarded as the spatial dimension $d$. 
However, the critical dimension is found to be $d=M+1=3$ and is not $d=M+1=2$, which should
naturally be expected by the one-parameter scaling hypothesis.
\end{abstract}



\maketitle


{\it Introduction-}
Since the proposal of Anderson, the localization of electron in disordered lattices has 
been one of the most fundamental problems associated with the essence of electron conduction 
process \cite{anderson58,lifshiz88,abrahams10}. 
No matter how high the spatial dimension may be, the Anderson localized state exist
prior to the delocalized conducting state, and a transition from localized state to the delocalized
state, the so called Anderson transition (AT), 
occurs as the relative strength of disorder decreases \cite{markos06,garcia07,garcia08,slevin14,tarquini17}. 
Theoretical predictions have been obtained by using several theoretical tools such as 
the one-parameter scaling hypothesis, the self-consistent theory, and so on
\cite{abrahams79,vollhard80}.  

On the other hand, in the study of chaotic systems the ergodic
transition of quantum maps is equivalent to the AT 
of disordered lattice \cite{casati79,casati89,borgonovi97}.
Upon this equivalence, the dynamical AT has been first experimentally
observed for the quantum map systems implemented on 
the optical lattice \cite{chabe08,lemarie10}. 
In this case the number of dynamical degrees of freedom corresponds to 
the number of spatial dimension of the disordered lattices, 
and so the features of AT in high-dimensional 
lattices can be explored by the quantum maps. 

The dynamical interaction among the degrees of freedom thus enables the 
delocalization transition. Then the following question immediately follows: 
can the Anderson loalization in the disordered lattices be destroyed 
as it is perturbed by dynamical degrees of freedom such as phonon modes?
The perturbation by infinitely many phonon modes can be modeled
by a stochastic perturbation, and it is well-known that the stochastic perturbation
destroys the localization and induces a normal diffusion \cite{haken72,palenberg00,moix13,knap17}. 
However, the effect of dynamical perturbation composed 
of {\it finite number} of coherent modes has still been unanswered.  
In the previous papers, we investigated the effect of
finite-number harmonic perturbations on one-dimensional disordered lattice
(ODDL), and showed that the ODDL exhibits a normal diffusion at least on a 
finite time scale \cite{yamada93,yamada98,yamada99}. 
On the other hand, numerical and mathematical
studies claim that the localization is persistent for finite-number harmonic perturbations
Refs.\cite{hatami16,bourgain04},
and which of localization and delocalization dominates has still been open to question.
It is quite interesting whether or not a coherent dynamical perturbation composed of
finite number of harmonic modes can dynamically destroy the localization.
In this letter, we present novel results replying the question.

{\it Model-}
We consider tightly binding ODDL perturbed by coherent periodic perturbations 
with different incommensurate frequencies. It is given by
\beq
\label{eq:model}
i \hbar \frac{\partial \Psi_n(t)}{\partial t}  &=& \Psi_{n-1}(t)+\Psi_{n+1}(t)+V_L(n,t)\Psi_n(t), 
\eeq
where $V_L(n,t)=V(n)[1+f(t)] $. The coherently time-dependent part $f(t)$ is given as,
 \beq
  f(t)=\frac{\eps}{\sqrt{M}} \sum_i^M\cos(\omega_i t), 
\eeq
where $M$ and $\eps$ are the number of frequency components and 
the strength of the perturbation, respectively.
Note that the long-time average of the total power of the perturbation is normalized to 
$\overline{f(t)^2}=\eps^2/2$.
The frequencies $\{ \omega_i\}(i=1,...,M)$ are taken as mutually incommensurate numbers of order $O(1)$.
The static on-site disorder potential takes random value $V(n)$ uniformly distributed 
over the range $[-W/2, W/2]$, where $W$ denotes the disorder strength.

It is important to note that the harmonic source can be interpreted as the quantum 
linear oscillator of the Hamiltonian $\sum_i^M\omega J_i$ interacting with the irregular lattice with the
quantum amplitude $\frac{\eps}{\sqrt{M}} \sum_i^M\cos \phi_i$ instead of the classical force
$f(t)$, where $(J_i,\phi_i)=(-i\pr/\pr\phi_i,\phi_i)$ are the action-angle
operators of the $i$-th oscillator.
Each quantum oscillator has the action eigenstates $|n_i>$ with the action eigenvalue $J_i=n_i\hbar~(n_i:$integer) 
and the energy $n_i\hbar\omega_i$. Thus the system (\ref{eq:model}) is 
regarded as a quantum autonomous system of $(M+1)$-degrees of freedom 
spanned by the quantum states $|n>\prod_{i=1}^M |n_i>$ \cite{hatami16}.

We take a lattice-site eigenstate as the initial state $|\Psi(t=0)\>$, 
i.e. $\<n|\Psi(t=0)\>=\delta_{n,n_0}$, 
 and numerically observe the spread 
of the wavepacket measured by the mean square displacement (MSD), 
$m_2(t) = \sum_{n}(n-n_0)^2 \left< |\Psi(n,t)|^2 \right>$.

First, we consider the limit $M\to\infty$. In this case $f(t)$ can be identified with the 
delta-correlated stochastic force $<f(t)f(t^{`})>=\Gamma \delta(t-t^{`})$, 
where $\Gamma \propto \eps^2$ is a noise strength.
The localization is surely destroyed and the normal diffusion $m_2(t)=Dt$ with the diffusion constant
$D$ is recovered for $t\to \infty$ \cite{yamada98,yamada99}, 
as was first pointed out by Haken and 
his coworkers \cite{haken72,palenberg00}.  
They predicted analytically for the white Gaussian noise 
\beq
D=\lim_{t \to \infty}\dfrac{m_2(t)}{t} \propto \frac{\Gamma}{\Gamma^2+W^2/12},
\label{eq:D-stochastic}
\eeq
for weak enough $\eps$. 
If $W\gg \Gamma$, $D\propto W^{-2}$ but recently
it was shown that $D\propto W^{-4}$ 
for strong disorder region $W\gg 1$ \cite{moix13}.
The noise-induced diffusion has been extended for a random 
lattice driven by the colored noise, including 
the fluctuation of the off-diagonal terms \cite{palenberg00,moix13,knap17}. 

However, for finite $M$, $f(t)$ can no longer be replaced by the random noise, and it plays as a
coherent dynamical perturbation, and the system is a quantum dynamical system
with $(M+1)$-degrees of freedom. 
The main purpose of this study is to investigate how does 
the nature of the quantum dynamics of the irregular lattice changes as the number $M$ decreases
from $\infty$ to $0$.

\begin{figure}[htbp]
\begin{center}
\includegraphics[width=8.7cm]{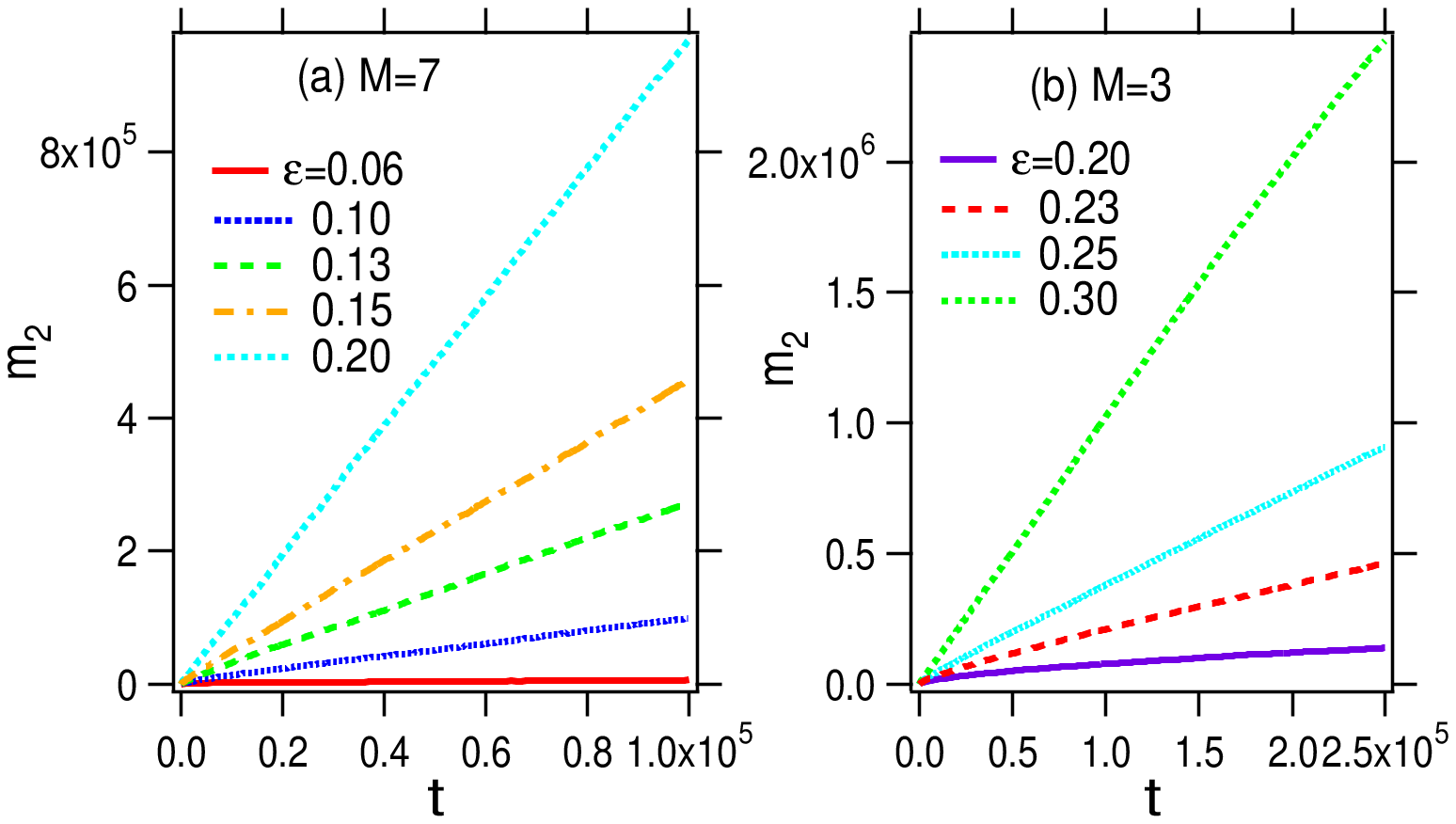}  
\hspace{7mm}
\includegraphics[width=4.4cm]{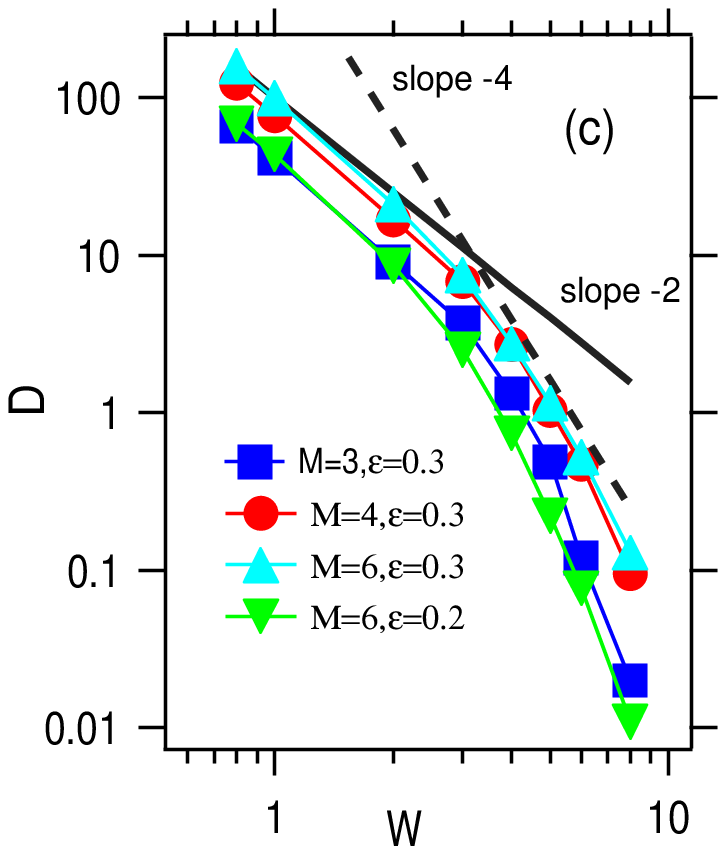}  
\hspace{-7.2mm}
\includegraphics[width=4.4cm]{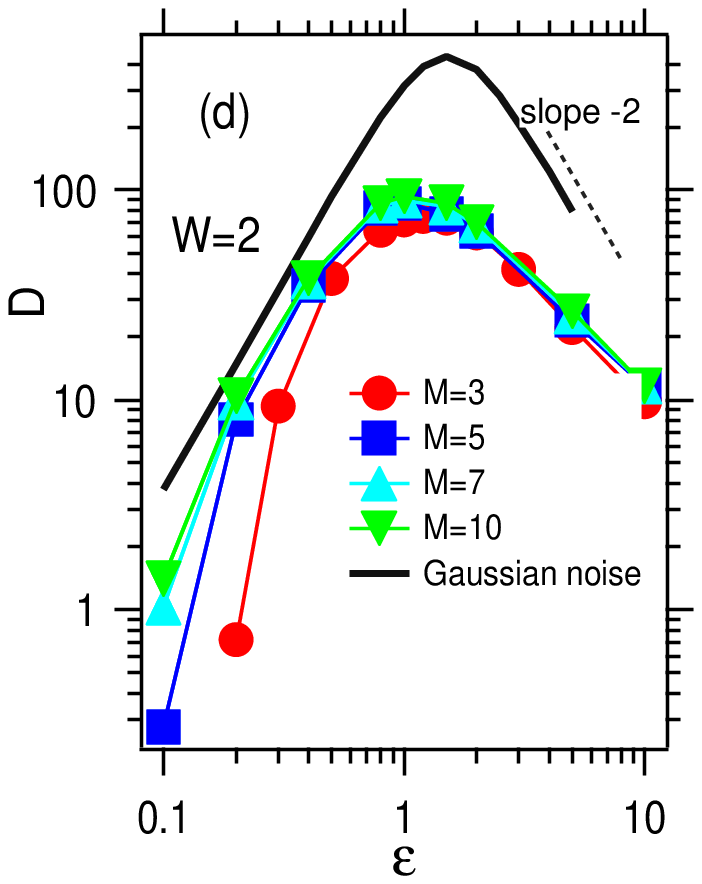}  
\caption{(Color online)
The $m_2(t)$ as a function of time in the ODDL of 
(a)$M=7$ and (b)$M=3$ with $W=2$ for 
some values of the perturbation strength $\eps$ 
 increasing from $\eps=0.06$(bottom) to $\eps=0.2$(top) for $M=7$
and from $\eps=0.2$(bottom) to $\eps=0.3$(top) for $M=3$, respectively. 
Note that the axes are in the real scale.
(c)The diffusion coefficient $D$ as a function of $W$ 
and (d) the $D$ as a function of $\eps$ for several $M$, 
determined by the least-square-fit for the $m_2(t)$ for $t>>1$.
The system and ensemble sizes are 
$N=2^{14} \sim 2^{15}$ and $10 \sim 40$, respectively,  
throughout this paper.
We used 2nd order symplectic integrator
with time step size $\Delta t=0.02 \sim 0.05$, and 
take $\hbar=0.125$ as the Planck constant.
}
\label{fig:dif-con-W}
\end{center}
\end{figure}

{\it Delocalized states and normal diffusion-}
We show typical examples of time evolution of MSD for $M=7$ and $M=3$ 
in Fig.\ref{fig:dif-con-W}(a) and (b), respectively. 
If $\eps$ is large enough, it is evident that MSD follows asymptotically
the normal diffusion $m_2=Dt$, which means that only a finite number of 
coherent periodic modes plays the same role as the stochastic perturbation
in the disordered lattice. The $W-$ and $\eps-$dependence 
of the diffusion constant $D$ depicted
in Fig.\ref{fig:dif-con-W} follow the main feature of 
the stochastically induced diffusion constants:
as shown in the Fig.\ref{fig:dif-con-W}(c) the $W-$dependence  
changes from $D \propto W^{-2}$ for weak $W$ in Eq.(\ref{eq:D-stochastic})
 to $D\propto W^{-4}$ for $W \gg 1$, 
 following the theoretical prediction of stochastic perturbations \cite{moix13}.
Moreover, as depicted in the Fig.\ref{fig:dif-con-W}(d), even for $M=3$
the $\eps-$dependence reproduces the characteristic behavior of
the stochastically induced $D$, which first increases but finally decreases with $\eps$
after reaching to a maximum value. It is a remarkable feature of 
ODDL that a normal diffusion, which mimics the one induced by a stochastic force composed
of infinite number of frequencies, is self-generated by a coherent perturbation 
composed of only three incommensurate frequencies.

\begin{figure}[htbp]
\begin{center}
\includegraphics[width=7.0cm]{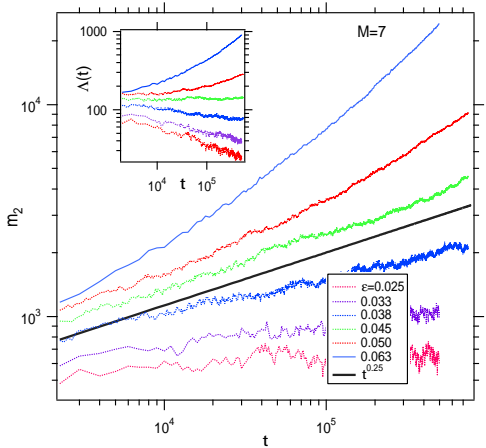}
\caption{(Color online)
The double-logarithmic plots of 
$m_2(t)$as a function of time
for some values of the perturbation strength $\eps$ 
 increasing from $\eps=0.025$(bottom) to $\eps=0.063$(top), 
where the diffusion exponent $\alpha$ is determined by 
the least-square-fit for the $m_2(t)$ with the critical case, 
in the perturbed ODDL of $M=7$ with $W=2$.
The data near the critical value $\eps_c$ are shown by bold black lines.
$\eps_c\simeq 0.045$, $\alpha \simeq 2/8=0.25$.
Note that the axes are in the logarithmic scale.
The inset shows the scaled MSD $\Lambda(t)$.
}
\label{fig:c3-s1s3-msd}
\end{center}
\end{figure}

On the other hand, the coherently perturbed ODDL always undergoes a definite
phase transition from the diffusing state to a localized state as $\eps$ decreases
crossing over a critical value $\eps_c$. 
The transition is quite similar to the AT of high-dimensional 
disordered lattice. 
As shown in Fig.\ref{fig:c3-s1s3-msd}, at $\eps=\eps_c$, the MSD exhibits
a subdiffusion $m_2 \sim t^\alpha$ with a critical diffusion index $\alpha~(0<\alpha<1)$.
Close to $\eps_c$, typical critical transient phenomena are observed. To show them we define 
the function $\Lambda(t)$ as the scaled MSD:
\beq
\Lambda(t)\equiv \frac{m_2(t)}{t^{\alpha}}, 
\eeq
divided by the subdiffusion. In the inset of the Fig.\ref{fig:c3-s1s3-msd} the $\Lambda(t)$ 
at various $\eps$ close to 
$\eps_c$ are displayed for $M=7$, which form a characteristic fan pattern spreading outward.

\begin{figure}[htbp]
\begin{center}
\includegraphics[width=4.2cm]{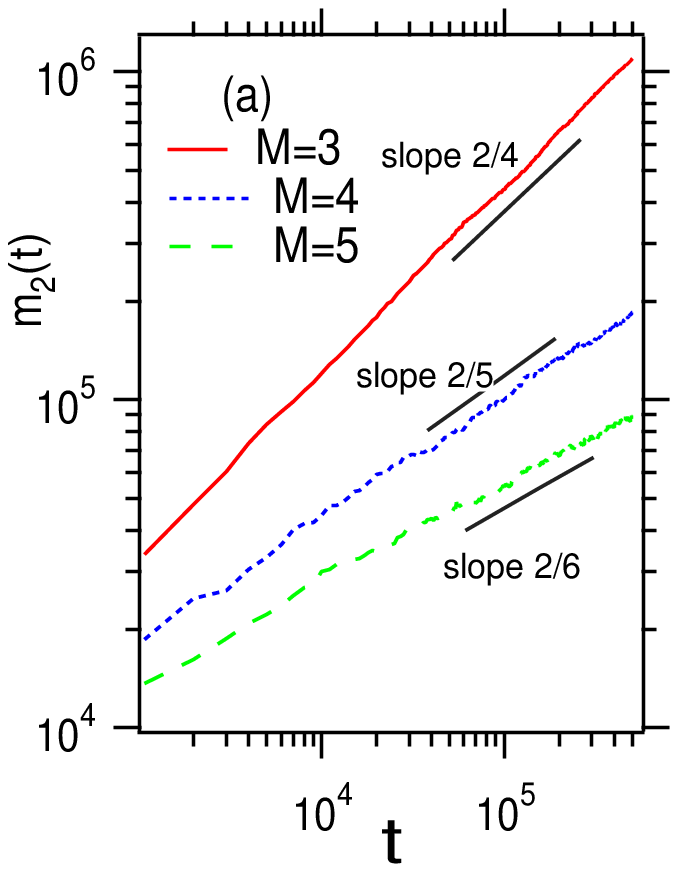}
\hspace{-0.5mm}
\includegraphics[width=3.9cm]{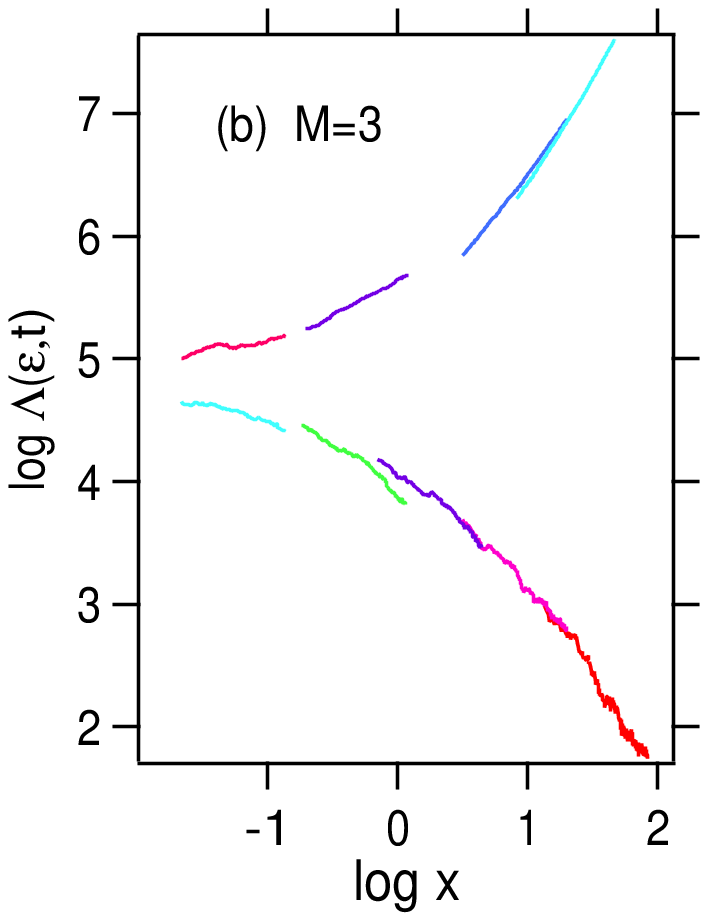}
\hspace{9mm}
\includegraphics[width=3.9cm]{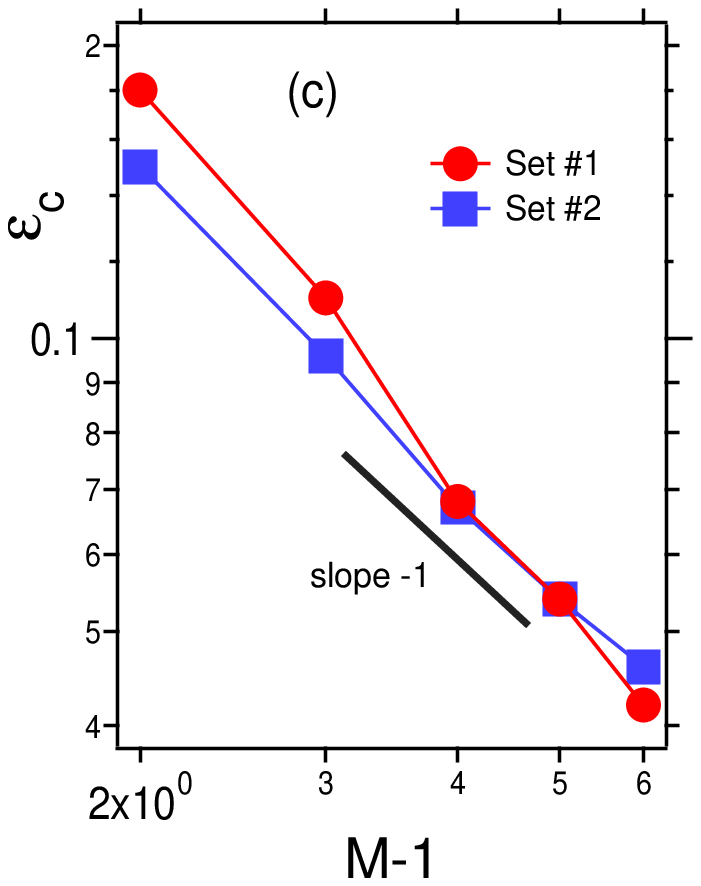}
\hspace{-0.8mm}
\includegraphics[width=4.0cm]{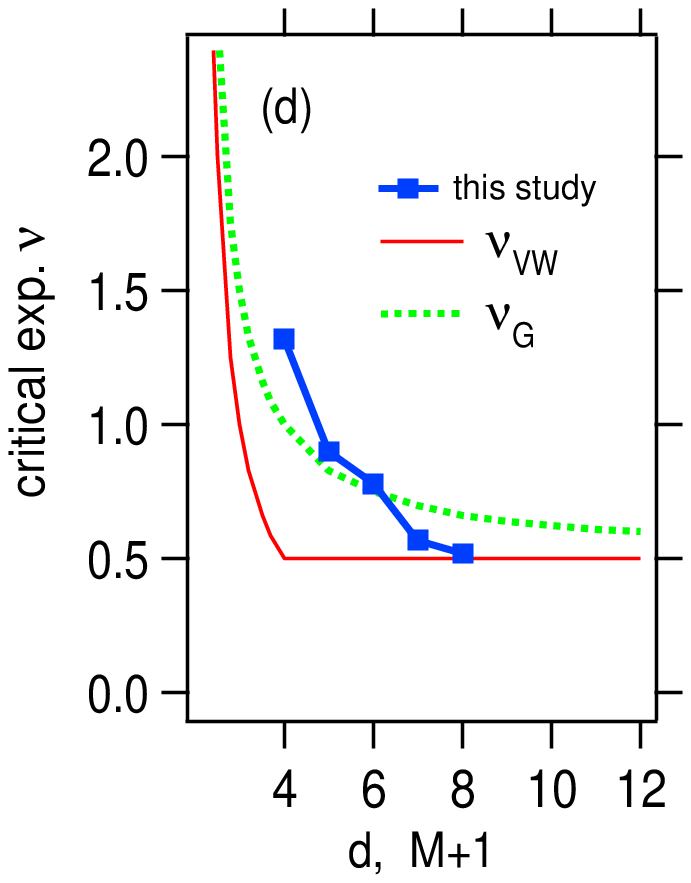}
\caption{(Color online)
(a)The double-logarithmic plots of $m_2(t)$ 
as a function of time near the critical pints $\eps_c$ in
the polychromatically perturbed ODDL with $W=2$  ($M=3,4,5$ from top).
(b)The scaled variable $\log \Lambda(\eps,t)$ 
as a function of $x=\xi_M(\eps)t^{\alpha/2\nu}$.
The delocalized(localized) regime is upper(lower) branch. 
(c)The critical perturbation strength $\eps_c$ as a function of $(M-1)$.
The result for different frequency set $\{ \omega_i \}$ is also entered.
Note that the axes are in the logarithmic scale.
The line with slope $-1$ is shown as a reference.
(d)The dimensionality $(M+1)=d$ dependence of the critical exponent $\nu$
which characterizes the critical dynamics. 
The  red solid line and green dashed line are 
 the results of the analytical prediction by 
$\nu_{VW}$ and $\nu_{G}$, respectively.
}
\label{fig:M3M4M5-epsc}
\end{center}
\end{figure}

As are demonstrated in Fig.\ref{fig:M3M4M5-epsc}(a), the index of
the critical subdiffusion decreases with $M$, following
the result of one-parameter scaling hypothesis
\beq
  \alpha = \frac{2}{d} = \frac{2}{M+1}
\label{eq:alpha-M}
\eeq
for the $d$-dimensional disordered lattice, if we regard $d$ as
the total number of degrees of freedom of our system, ie,. $d=M+1$,
which seems to be quite reasonable.

The localization in the side of $\eps<\eps_c$ is characterized by the localization
length $\xi_M$, which diverges at $\eps_c$ as $\xi_M(\eps) \sim (\eps-\eps_c)^{-\nu}$ with
the critical exponent $\nu(>0)$. A remarkable feature of the critical state is that, 
the fan pattern of $\Lambda(t)$ are represented by two unified curves depending whether $\eps>\eps_c$
or $\eps<\eps_c$ by using the scaling variable $x=\xi_M(\eps)t^{\alpha/2\nu}$, as demonstrated
by Fig.\ref{fig:M3M4M5-epsc}(b) for $M=3$. The $d=M+1-$dependence of the critical index $\nu$ is
shown in Fig.\ref{fig:M3M4M5-epsc}(d).
The more details of the finite-time scaling analysis for the numerical data 
are given in Refs.\cite{lemarie10} and \cite{yamada20}. 

Such a remarkable critical subdiffusion exists at $\eps_c$ for an arbitrary $M$, 
but the critical value $\eps_c$ decreases with $M$:
\beq
\eps_c \propto \frac{1}{(M-1)^\delta} , ~~~~~~~~~~\delta \simeq 1,
\label{eq:epsc-M}
\eeq
which does not depend upon $W$ as shown in Fig.\ref{fig:M3M4M5-epsc}(c). 
Thus the ODDL is always
localized if $\eps$ is small enough, but no matter how small $\eps$
may be, a normal diffusion mimicking a stochastically 
induced diffusion is realized if $M$ is taken large enough.

The mathematical research of \cite{bourgain04} for a very similar model to ours asserts 
the localized phase exists for small enough $\eps$ as long as $M$ is finite. 
In particular, the persistency of the localization for $M=2$ was numerically 
confirmed up to a value of $\eps$ beyond the perturbation regime \cite{hatami16}. 
On the other hand, 
in the large limit of $M$, the perturbation can be well approximated by a white noise, 
which makes the system delocalize for any $\eps \neq 0$ \cite{haken72,palenberg00,moix13,knap17}, 
To be compatible with the above observations,
 a  delocalization-localization transition (DLT) should 
exists for arbitrary large finite $M$, and it should disappear in a limit $M\to \infty$,
 which is just the background supporting our result of Eq.(\ref{eq:epsc-M}). 
An important fact is that the change 
to the delocalized state is not a crossover process but a quantum phase transition.

It is quite interesting that the dependencies of both $\alpha$
and $\eps_c$ upon $M$ are the same as those of the AT
observed for the quantum maps simulating the high-dimensional disordered lattice 
\cite{yamada15,yamada18,yamada20}.
If we are allowed to extrapolate the above results
for the smaller $M$, $\eps_c$ diverges at $M=1$,
at which the critical diffusion index becomes $\alpha=1$.
This fact implies that for $M=1$ the critical subdiffusion
is realized at $\eps=\eps_c=\infty$ as a normal diffusion; namely, 
that $M=1$ is the critical dimension of the DLT, 
which has been established for the quantum maps and high-dimensional 
disordered lattices. 
However, our numerical results reject the above conjecture.

\begin{figure}[htbp]
\begin{center}
\includegraphics[width=9.0cm]{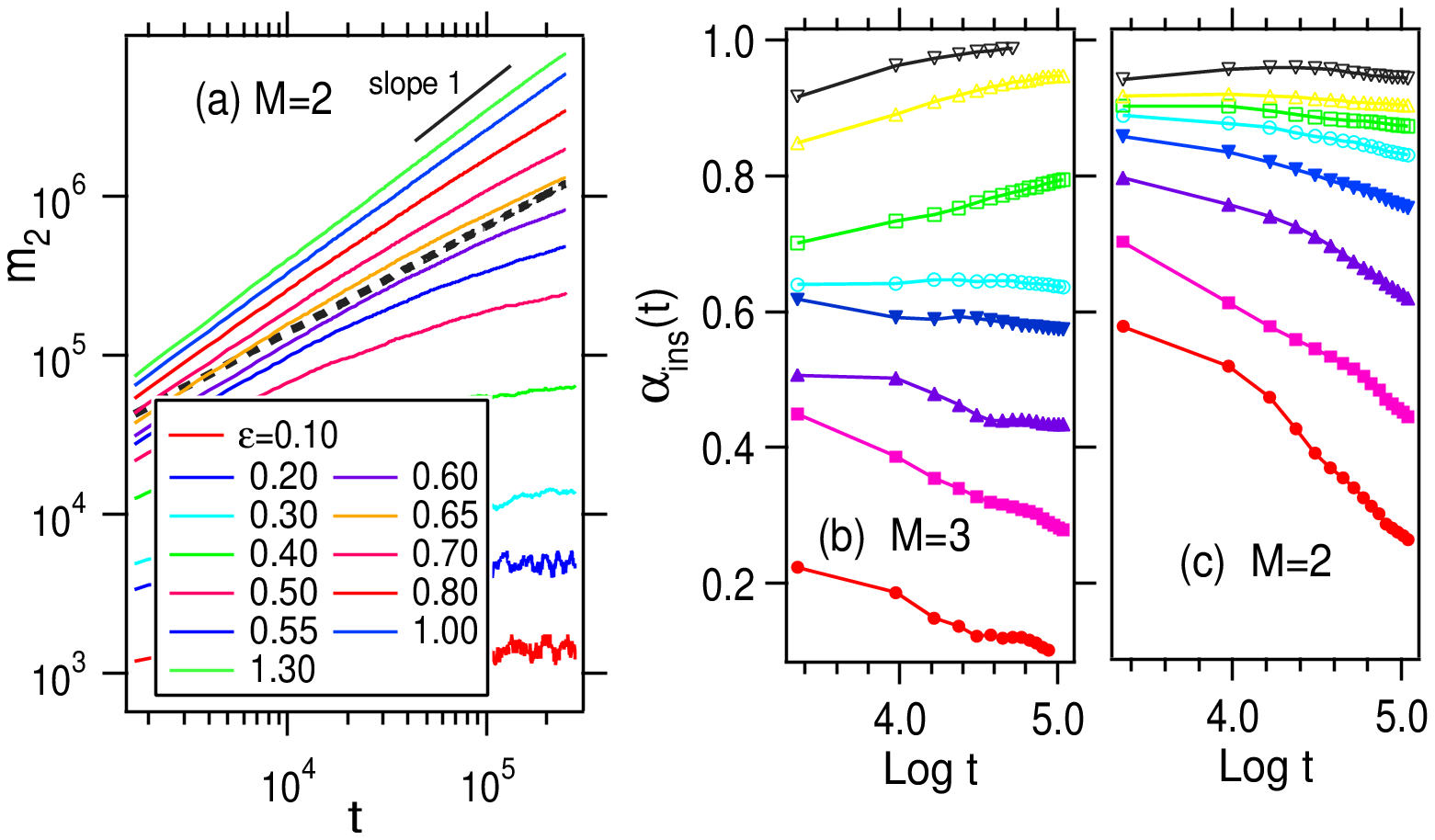}
\hspace{4mm}
\includegraphics[width=5.5cm]{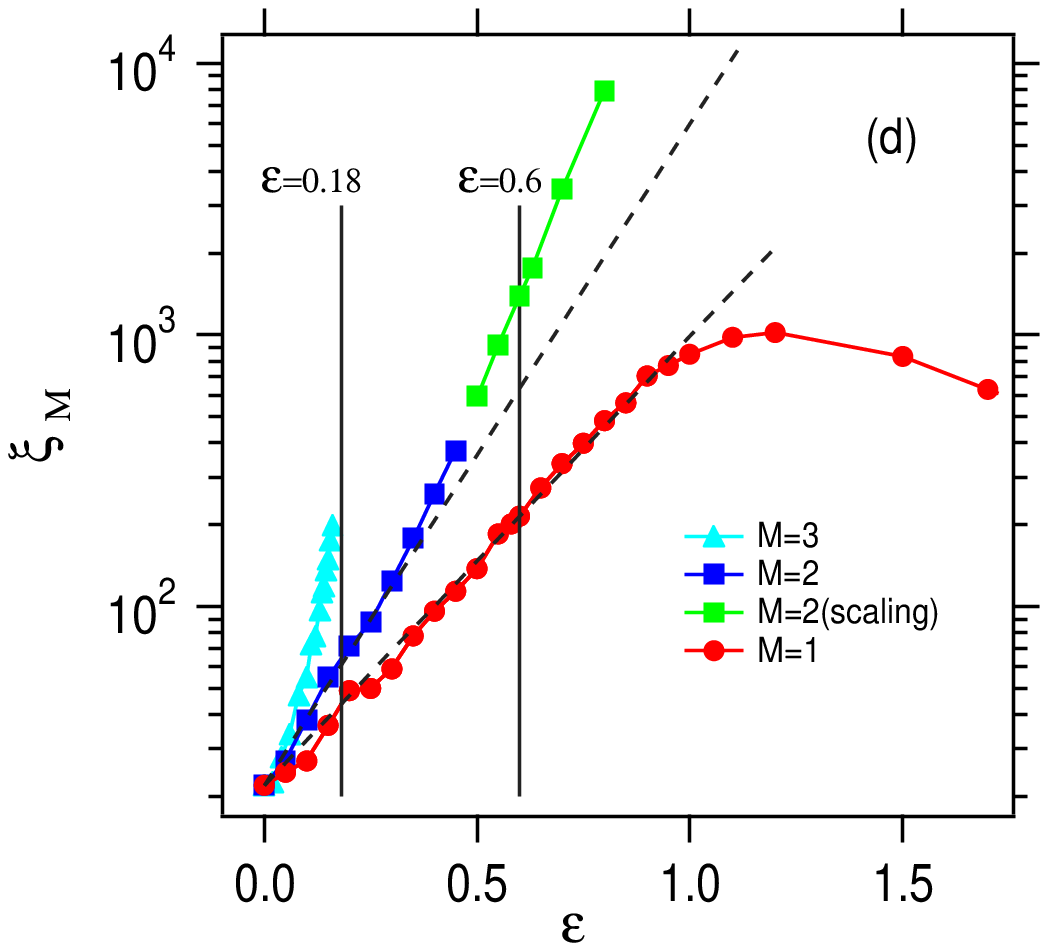}
\caption{(Color online)
(a)The double-logarithmic plots of $m_2(t)$
as a function of time for some values of the perturbation strength $\eps$ 
 increasing from bottom to top
in the trichromatically perturbed ODDL of $M=2$.
The panels (b) and (c) are the diffusion exponent $\alpha_{ins}(t)$ 
as a function of time in the cases $M=3$ and $M=2$, respectively.
(d)Localization length as a function of $\eps$ for
$M=1,2,3$ with $W=2$.
Some LL of $M=2$ are obtained by scaling relation 
$m_2(t) \sim \xi_M(\eps) F(t/\xi_M(\eps)^2)$ for $\eps \geq 0.5$.
Note that the horizontal axis is in logarithmic scale.
The dashed lines show $\xi_M \propto e^{5.5\eps}$ 
and $\xi_M \propto e^{3.8\eps}$, respectively.
The lines $\eps=0.18$ and $\eps=0.6$ are shown as a reference,
and $\xi_M(\eps=0)\simeq 20$ for $W=2$}.
\label{fig:c3c2-msd-alpha}
\end{center}
\end{figure}

{\it Number of critical modes ($M=2$)-}
If the above conjecture is correct,
$M=2~(d=3)$ should exhibit the critical 
phenomenon. In Fig.\ref{fig:c3c2-msd-alpha}(a) the log-log plot of MSD curves for $M=2$
are shown for various values of $\eps$. 
Surely, the $m_2(t)$ of $\eps=0.6$, which
roughly predicted from the interpolation of the numerical data for $M\geq 3$ 
follows the expected critical subdiffusion of the exponent $\alpha=2/3$
predicted by Eq.(\ref{eq:alpha-M}) in the
initial stage, but it drops from the straight line as the 
time elapses.

To overview the whole feature of the MSD curves, it is instructive to show
the time evolution of the diffusion exponent defined as the
instantaneous slope of the log-log plot of MSD 
\beq
\alpha_{ins}(t)=\dfrac{d\log m_2(t)}{d\log t}.
\eeq
If DLT happens at a finite $\eps=\eps_c$, then the $\alpha_{ins}(t)$ should keep a constant value
$\alpha(\eps_c)<1$.
Above $\eps_c$, as time passes, $\alpha_{ins}(t)$ increases up to the exponent 
$1$ indicating the normal diffusion, while it decreases to zero indicating the localization
 below $\eps_c$. Indeed, the expected feature is evident for the $\alpha_{ins}(t)$ plot of $M=3$ 
shown in Fig.\ref{fig:c3c2-msd-alpha}(b)
The same feature is observed also for $M\geq 4$.

However, as shown in Fig.\ref{fig:c3c2-msd-alpha}(c)
the $\alpha_{ins}(t)$-plots of $M=2$
shows a quite different feature. No curves follow the critical behavior 
$\alpha_{ins}(t)=$const$<1$, and all the curves tends to decrease from the
initial values, which approaches to $1$ as $\eps$ increases. 
As $\alpha_{ins}(t)$ comes close to 1, the time scale beyond which $\alpha_{ins}(t)$ 
begins to decrease becomes longer.
Certainly it seems as if the normal diffusion 
$\alpha_{ins}(t)=1$, which would be realized in the limit $\eps\to\infty$,
were the critical diffusion. These facts indicates that the DLT does not exists for $M=2$
in contradiction with the prediction of the Eqs.(\ref{eq:alpha-M}) and (\ref{eq:epsc-M}), 
 and that $M=2(d=3)$ is the critical dimension.

{\it Comparison by localization length-}
Localization, of course, occurs with $M=1$. Then what is the difference of the localizations 
between the case of $M=1~(d=2)$ and the case of $M=2~(d=3)$. 
In both cases of $M=d-1=1$ and $M=d-1=2$, the localization length grow
 exponentially when the $\eps$ is small enough ($\eps<0.8$ for $M=1$  
 and $\eps<0.3$ for $M=2$), which coincides with the case of the $d=2$ ordinary 
disordered lattice.

However with a further increase of $\eps$, $\xi_M$ begins to decrease steeply for $M=1$.
Such a behavior is a direct reflection that the inter-site transfer is suppressed 
by the random potential enhanced with the increasing perturbation strength $\eps$.
Let us remember that as shown in Fig.1(d) even the recovered diffusion constant of the
the system of $d = M+1 \gg 1$ in general decreases steeply with $\eps(>\eps_c)$.
This is the reason why, unlike the ordinary $d-$dimensional irregular lattice, 
$d=2$ can not be the critical dimension of our system.   
The growth of $\xi_M$ with $\eps$ takes place only by increasing the dimension from
$d=2$ to 3. 

Indeed, for $d=M+1=3$ the localization still remains, but $\xi_M$ increases with $\eps$
exponentially. The exponential growth rate is further enhanced and a super-exponential growth 
occurs as $\eps$ increases beyond $~O(1)$, as is depicted in Fig.\ref{fig:c3c2-msd-alpha}(d).
And it is for $M=3$ that the divergence of $\xi_M$ is first observed at a finite $\eps_c$.

{\it Summary and discussion-}
In the present paper, we investigated the delocalized and the localized
motion in 1D irregular lattice coherently perturbed by the harmonic modes.
In order to induce a delocalized motion the stochastic perturbation
composed of infinite number of harmonic mode is not necessary: 
the diffusive motion is always induced only by a few number of harmonic modes
if the perturbation strength is strong enough. 
The critical perturbation strength ($\eps_c$) and the critical subdiffusion exponent
($\alpha$) decrease with the number of modes $M$, and their dependencies upon
$M$ are almost same as those of the Anderson transitions 
numerically established for the multi-dimensional quantum maps,
 which can be considered
as modified versions of the many-dimensional Anderson model \cite{yamada20}.
However, the critical number of the degrees 
of freedom is not $d=M+1=2$ but $d=3$ in our system. 
Thus our system provides with an example 
demonstrating that the critical dimension of the DLT may be larger than $d=2$ and  
depend upon the nature of recovered diffusion,
as summarized in the Table \ref{table1}.

The Anderson-like transition discussed in the present paper affords a new example
of quantum phase transition with which the coherent localized state changes to 
a decoherent diffusive state.
Existence of such a kind of quantum transition have been known in some quantum chaos 
systems which exhibits chaotic diffusion \cite{casati89,chabe08}. More generally, it 
will play a crucial role when quantum systems of small number of degrees of freedom 
get ergodic properties \cite{matsui16b}.  We expect that investigations
for the transition process to the decoherent
and delocalized states in the quantum dynamical systems with small degrees of freedom 
would contribute much to the fundamental study of quantum statistical physics.

\begin{table}[htbp]
\begin{center}
 \caption{
Dimensionality of the DLT.
For $4 \leq M <\infty$ the result is same as the case of $M=3$.
The lower lines is result of the $d-$dimensional disordered systems.
Loc: exponential localization, Diff:Normal diffusion.
\label{table1}
}
 \begin{tabular}{lccccccc}
\hline
\hline
$d(=M+1)$ &1 & 2 & 3 & 4 & 5 &...& $\infty$ \\ \hline 
this study  & Loc & Loc & Loc & DLT & DLT &...& Diff\\ 
quantum maps \cite{yamada20}  & Loc & Loc & DLT & DLT  &DLT &...&Diff\\ 
Anderson model    & Loc & Loc & DLT & DLT & DLT &...& DLT\\ \hline
\hline 
 \end{tabular}
\end{center}
 \end{table} 



\end{document}